\documentclass[epj-spec]{svjour}
\usepackage{graphics}
\usepackage{epsfig}
\begin{document}

\title{Local influence of boundary conditions on a confined
supercooled colloidal liquid}

\author{Kazem V. Edmond \and Carolyn R. Nugent \and Eric R. Weeks\thanks{\email{erweeks@emory.edu}}}
\institute{Physics Dept., Emory University, Atlanta, GA 30322, USA}

\abstract{
We study confined colloidal suspensions as a model system which
approximates the behavior of confined small molecule glass-formers.
Dense colloidal suspensions become glassier when confined between
parallel glass plates.  We use confocal microscopy to study the
motion of confined colloidal particles.  In particular, we examine
the influence particles stuck to the glass plates have on nearby
free particles.  Confinement appears to be the primary influence
slowing free particle motion, and proximity to stuck particles
causes a secondary reduction in the mobility of free particles.
Overall, particle mobility is fairly constant across the width of
the sample chamber, but a strong asymmetry in boundary conditions
results in a slight gradient of particle mobility.
}

\maketitle

\section{Introduction}

Glasses can form by rapidly cooling some liquids, or putting
them under large pressures \cite{angell00}.  While this process
of glass formation has been studied for years, questions still
remain \cite{angell00q}.  One method of study has been to examine
the properties of glass forming materials when they are confined.
This confinement can be in thin films, nanocapillary tubes, or
nanoporous materials \cite{mckenna05,roth05}.  In some cases
confined materials become more glassy, while in other cases
they become less glassy.  A key factor seems to be the boundary
conditions \cite{mckenna05,kob02,lowen99,ngai02}.  ``Sticky''
boundary conditions, where the sample forms strong chemical
bonds to a wall, tend to result in glassier behavior.  ``Free''
boundary conditions, such as a free-standing polymer thin film, tend
to result in more liquid-like behavior.

Colloidal suspensions are composed of small particles in a liquid,
with particles of diameter a few microns or smaller.
These are a good model system to study the glass transition
\cite{pusey86,weeks00,brambilla09,snook91,vanmegen98,courtland03,lynch08}.
The transition has one main control parameter,
the volume fraction $\phi$ of the solid particles.
When $\phi > \phi_g \approx 0.58$, the sample appears
glassy.  Macroscopically, such samples do not flow, and
microscopically, particles no longer diffuse through the sample
\cite{brambilla09,snook91,vanmegen98,courtland03,lynch08,cheng02}.
Previous experiments showed that confinement results in
slower colloidal dynamics, far more than would be due to
hydrodynamic drag near a wall \cite{nugent07prl,sarangapani08}.
These experiments used either smooth glass walls, or walls with
a few colloidal particles stuck to them.  It was unclear if the
boundary conditions made a strong difference, as in all cases the
confined particles diffused more slowly, and the studies did not
look for differences based on boundary conditions.

In this work, we reexamine the data of Ref.~\cite{nugent07prl},
studying how local variations in the boundary conditions influence
particle motion.  We find that mobile particles which are close
to immobile particles (ones stuck to the confining sample chamber
walls) have slower motion.  This also influences how particles
rearrange; rearrangements are easier farther away from stuck
particles.  Our results show that variability in the boundary
conditions exerts a localized influence on particle motion,
enhancing the glassiness of the confined sample.

\section{Experimental Methods}

We use colloidal poly-methyl-methacrylate (PMMA) particles for
our samples.  These particles are sterically stabilized with a
short polymer layer of poly-12-hydroxystearic acid
(PHSA), which protects
them from aggregating \cite{antl86}.  In certain solvents
these particles behave as good approximations to hard spheres,
with interactions only when they are close enough for the PHSA
polymers to overlap \cite{pusey86,brambilla09,snook91,vanmegen98}.
Our solvent mixture is composed of cyclohexylbromide and decalin,
chosen to match the index of refraction and the density of the PMMA
particles \cite{dinsmore01}.  In this mixture, the particles acquire
a slight charge \cite{dinsmore01}.  (This charge is most noticeably
manifested as a shifting of the first peak of the pair correlation
function to positions slightly larger than the particle diameter, which would
be the position for hard spheres.)  To frustrate crystallization
in our samples, we use a mixture of two particle sizes with mean
radii $a_S = 1.18$~$\mu$m and $a_L = 1.55$~$\mu$m, and with
individual volume fractions $\phi_S = 0.26$, $\phi_L = 0.16$,
and $\phi_{\rm tot}=0.42$ (sample A in Ref.~\cite{nugent07prl}).
The mean particle radii have an uncertainty of $\pm 0.02$~$\mu$m
and a polydispersity of 5\% (samples of each individual species
eventually can form colloidal crystals, indicating that they are
not too polydisperse).

The particles are observed using confocal microscopy
\cite{prasad07}.  Our microscope (``VT-Eye,'' Visitech) can scan
two-dimensional (2D) images at a frame rate of 92 images/s, and
we scan a three-dimensional (3D) image stack composed of 100~2D
images once every 2.0~s, for the data presented in this work.
In our solvent mixture with viscosity $\eta = 2.25$~mPa$\cdot$s,
the time for a small particle to diffuse its own radius in a
dilute suspension is roughly 3 s.  In our concentrated samples,
diffusion is much slower, and so our 2.0~s interval between
image stacks is sufficiently rapid to follow the motion of all of
the particles in 3D using standard particle tracking techniques
\cite{dinsmore01,crocker96}.  Our image stacks are of dimensions
$50 \times 50 \times 20$~$\mu$m$^3$, and particle positions are
resolved to within 50~nm in $x$ and $y$ and 100 nm in $z$.  To aid
in image analysis, only the small particles are dyed, and thus we
cannot observe the behavior of the large particles using confocal
microscopy.  We have used differential interference contrast microscopy
(DIC) to verify that the large particles are behaving similarly
to the small particles, and other experiments \cite{lynch08}
and simulations \cite{desmond09} give us confidence that the two
particle species should not behave too differently in either their
positions or their dynamics.

We place these samples in wedge-shaped sample chambers, with
a wedge angle of less than a degree.  This allows us to study
the same sample at a variety of thicknesses $H$, down to $H
= 6$~$\mu$m.  We have
never seen an influence of the gradient in $H$ in any of our
data \cite{nugent07prl,edmond10}.  Further details of our sample
chambers are discussed in Ref.~\cite{edmond10}.

\begin{figure}
\centering
\includegraphics[width=11.0cm]{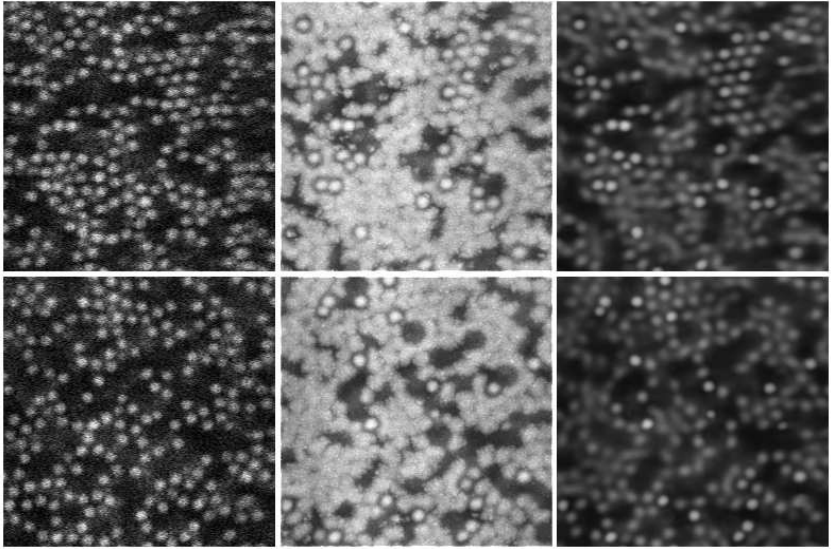}
\caption{
Images showing behavior near the top wall ($z=H$, top row of images)
and near the bottom wall ($z=0$, bottom row of images).  The left
images are individual 2D slices from a single 3D image stack.
The middle images show the maximum intensity observed at each
location over the duration of the movie (2300~s in duration).
The right images show the average intensity observed at each
location over the duration of the movie.  Bright spots in the middle
and right images indicate visible particles stuck to the glass,
and darker regions suggest where the larger invisible particles
are stuck.  Regions that are black in the left panels but not in
the middle panels are probably occupied by mobile large particles.
For this region of the sample chamber, $H =
8.06$~$\mu$m.
}
\label{montageone}
\end{figure}

Our sample chamber walls are flat, untreated glass 
(a glass coverslip on one side, a glass microscope slide on
the other).  Due to the slight charges on the particles and the
glass walls, the particles are repelled from
the walls.  However, we observe over long periods of time that
some particles become stuck to the glass, and it is the influence
of these stuck particles that we investigate in this article.
These particles are visible in Fig.~\ref{montageone}.  The top row
shows images corresponding to the top of the slide, and the bottom
row shows images from the bottom of the slide in the same $x$
and $y$ location.  The left images are 2D raw images taken from
the confocal movie.  The middle panels are the maximum intensity
observed at each pixel over the duration of the movie.  In these
images, stuck small particles show up as distinct white circles,
and stuck large particles show up as black regions.
The remaining space is filled
in fairly uniformly by the visible small particles exploring the
available free space.  The right panels of Fig.~\ref{montageone}
show the time-averaged intensity of each pixel.  Here the brightest
white circles are the stuck particles, and the less bright spots
are where mobile particles spent longer periods of time.  
The darker regions correspond to locations where no visible
particles wandered during our observation period, and thus are
likely to have invisible particles stuck in those locations.

The images of Fig.~\ref{montageone} show that the stuck particles
are easy to distinguish in the data, as they do not move.
We also note that their positions are closer to the wall than
the other particles, both giving us evidence for the repulsion
of the mobile particles from the walls, and also allowing us
to accurately determine the local thickness $H$ based on the
stuck particle positions \cite{edmond10}.  
The area fraction of the stuck particles ranges from 10\% to 20\%
(total for both species).  From DIC microscopy
as well as images such as those shown in Fig.~\ref{montageone}, we
find that the large particles appear to be stuck in comparable area
densities to the small particles.  It is interesting to also note
that there are large patches next to the walls which do not have
any stuck particles.  This can be see in the connected regions that
are uniformly bright in the middle images of
Fig.~\ref{montageone}.
While the avoided regions make it clear where the stuck particles
are located, the regions fully explored by the mobile particles
given a sense of where the particles see only the flat glass wall.

From the $z$ positions of the stuck particles, we can determine the
sample thickness $H$.  This is the maximum range that the particle
{\it centers} can cover, whereas the actual wall spacing is $H
+ 2 a_S$.  As we can determine $H$ more precisely than we know
$a_S$, we report our results in terms of $H$.  The uncertainty
in measurements of $H$ is $\pm 0.01$~$\mu$m and $2 a_S = 2.36 \pm
0.04$~$\mu$m \cite{nugent07prl}.  Our uncertainty of $H$ is low,
much smaller than our resolution (0.1~$\mu$m in $z$), because we
average over all of the particles stuck on each surface to find
the surface's location in $z$.  While individual particles'
$z$-positions vary because of noise, the average is well defined
in all cases.

\section{Results}

\begin{figure}
\centering
\includegraphics[width=8.5cm]{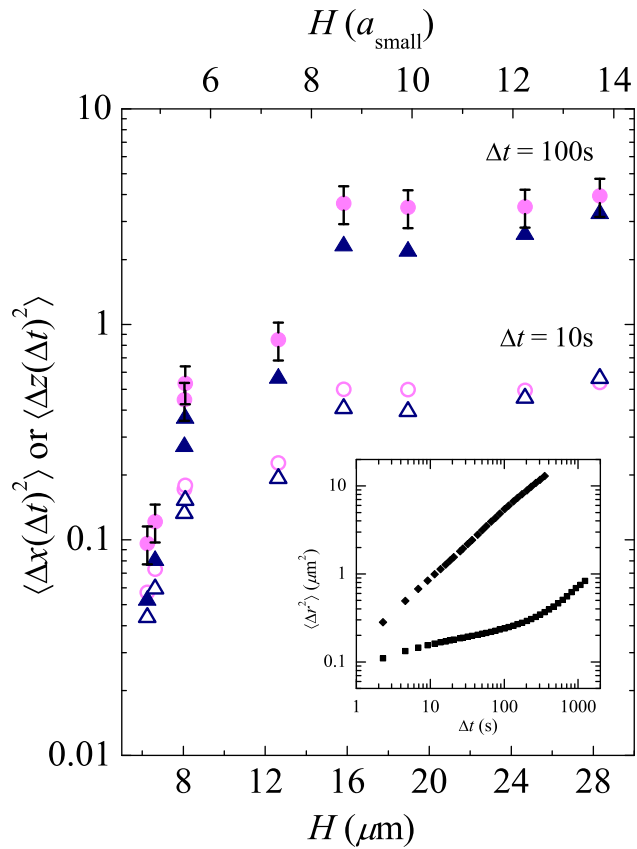}
\caption{
(Color online.)
The average mobility $\langle \Delta x^2 \rangle$ (parallel to the
sample chamber walls, indicated by circles) and $\langle \Delta
z^2 \rangle$ (perpendicular to the sample chamber walls, indicated
by triangles), as a function of the confinement thickness $H$.
The open symbols are for $\Delta t=10$~s and the solid symbols are
for $\Delta t = 100$~s.  
There is an additional data set at $H=69$~$\mu$m not shown which
agrees with the data shown for $H>16$~$\mu$m \cite{nugent07prl}.
The uncertainties of the data points are $\pm 20$\%, and are
indicated by representative error bars for one set of data.
The inset shows the mean square displacement for samples with
$H=16.32$~$\mu$m (upper curve) and $H=6.25$~$\mu$m (lower curve).
}
\label{avgmobility}
\end{figure}

As noted in the introduction, our prior work showed that confinement
results in slower particle motion.  This is illustrated in
Fig.~\ref{avgmobility}, which shows the mean distance particles
move $\langle \Delta x^2 \rangle$ (circles) and $\langle \Delta
z^2 \rangle$ (triangles) within a given time $\Delta t$, as a
function of the thickness $H$.  For $H > 16$~$\mu$m, there is no
dependence on $H$, suggesting that $H \approx 16$~$\mu$m is the
onset length scale for the confinement effects for this sample.
For smaller $H$, the motion slows down dramatically.  This suggests
that confinement induces the colloidal glass transition to occur
``earlier,'' that is, at a lower volume fraction $\phi$ than would
be seen in the bulk.  The results do not vary qualitatively with the time
scale $\Delta t$; in the results below, we use $\Delta
t = 100$~s as a representative time scale.  In our earlier work,
we found that the onset length scale grows with increasing
volume fraction as $\phi \rightarrow \phi_g$ \cite{nugent07prl}.
In this article we focus on a single sample with $\phi = 0.42$,
as discussed above.  (This sample is a liquid when unconfined,
and while it slows down upon confinement, the sample never enters
the glassy state for any confinement we have studied.)

\begin{figure}
\centering
\includegraphics[width=10.0cm]{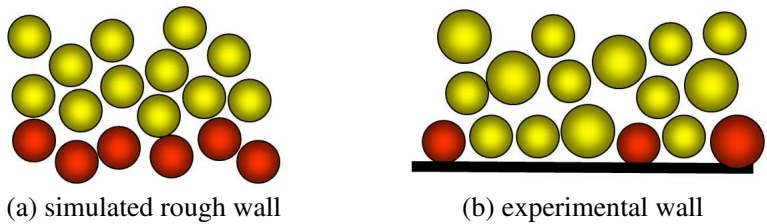}
\caption{
(Color online.)
Sketch of typical boundary conditions.  (a) Rough walls in a
simulation; the dark (red) particles are taken from a liquid
configuration and then immobilized.  (b)  Our boundaries with 
a few dark (red) particles stuck to the walls.
}
\label{roughness}
\end{figure}

Simulations found both faster and slower dynamics within
confined samples, depending on the texture of the
confining boundary.  For cases with smooth walls, particle motion
is faster, and for cases with rough walls, particle motion is
slower \cite{kob02,lowen99}.  To simulate rough walls, typically
one simulates a bulk liquid, and then locks the positions of
some of the particles into place to become the walls, as shown
in Fig.~\ref{roughness}(a).  Near the walls, this roughness
sterically frustrates particle motion parallel to the walls,
and thus intuitively it is sensible that these boundary
conditions result in glassier dynamics.  It has also been noted
that for smooth walls, particles tend to form layers near the walls
\cite{lowen99,desmond09,aim67}.  Particles which move within layers
then could potentially move more easily \cite{lowen99}.
In one set of simulations,
an extra spatially varying potential was added to either enhance or
prevent particle layering \cite{goel08}.  They found that particles
are indeed more mobile when organized into layers.  An implication
then is that in addition to the steric frustration of motion,
rough layers such as shown in Fig.~\ref{roughness}(a) prevent
particles from layering on the walls, and this will additionally
slow the dynamics.

Our boundary conditions, schematically shown in
Fig.~\ref{roughness}(b), present two conflicting possibilities.
Mobile particles are hindered in their motion when they
encounter stuck particles.  This should 
lead to glassier behavior when the samples are
confined (implying that more particles are close to the wall,
where they feel the influence of the stuck particles).
On the other hand, the flat patches allow particles to
form layers against the wall.  Furthermore, the stuck particles
are the same size as these layers, and so also help
formation of additional layers farther from the wall.
Indeed, we do see layering of the visible (small) particles, as
discussed in Ref.~\cite{nugent07prl,edmond10} (cf. dotted lines in
Fig.~\ref{nzmobile}).  Simulations suggest that the larger invisible
particles also form layers against the walls, although because of
the binary mixture, the layers of both particle sizes quickly wash
out away from the walls \cite{desmond09}.  Our prior work found
that particles prefer to move within these layers \cite{edmond10},
suggesting the possibility of enhanced mobility, based
on comparison with the simulations.  The one mitigating factor is
that particles within the layers next to the walls are a mixture of
the two particle sizes, and so the layers are more ``entangled''
than they would be if all of the particles were identical in
size \cite{lowen99,desmond09}.  It is worth noting that in our
prior work, we also studied cases with completely smooth walls
and no stuck particles.  In those experiments, confinement still
resulted in slower motion \cite{nugent07prl}.  This demonstrates
that the stuck particles are not solely responsible for confinement
induced slowing in our samples, although perhaps
they slow motion even further.

\begin{figure}
\centering
\includegraphics[width=8.0cm]{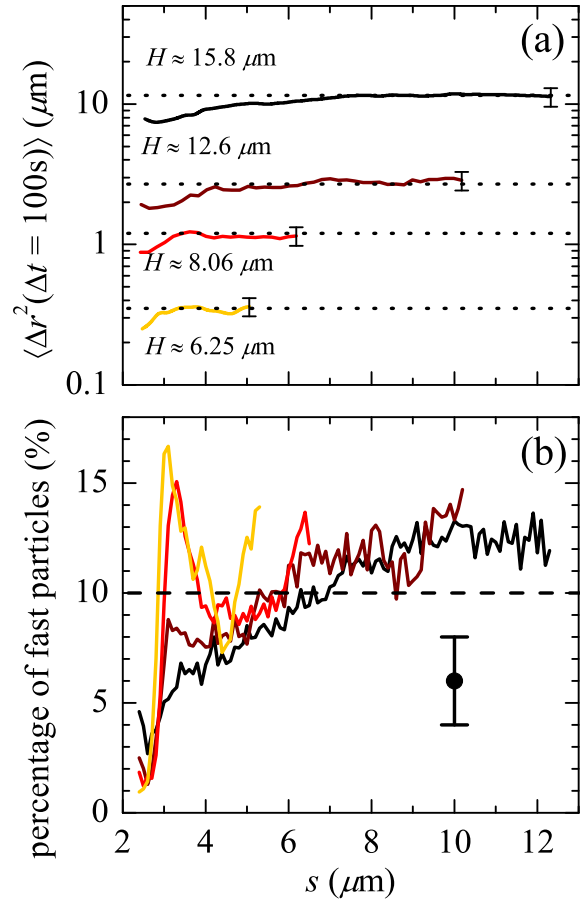}
\caption{
(Color online.)
(a) Graph of the mean square displacement $\langle \Delta r^2
\rangle$ as a function of the distance $s$ away from the nearest
immobile boundary particle.  The displacements are calculated
using a time lag $\Delta t = 100$~s, and the curves are from
different sample chamber thicknesses $H$ as indicated.  
The horizontal dotted lines indicate the plateau height for each
curve, averaged over all particles at all $s$.
The plateau height decreases for smaller $H$,
indicating the average slowing due to confinement.  Error bars
for each curve are indicated at the right end of each curve.
(b) Graph of the fraction of highly mobile particles, as a function
of distance $s$ from the nearest immobile boundary particle.  The
definition of ``highly mobile'' is such that 10\% of the
particles are considered highly mobile, so on average the data
should fluctuate around 10\% on this graph (indicated by the
horizontal dashed line).  The color/shading
of each curve is the same as for panel (a).  The error bar shown
indicates the uncertainty of the data.
In both panels (a) and (b), the
curves are
truncated at large $s$, where there are too few particles
to provide adequate
statistics.
}
\label{stuckmobility}
\end{figure}

To clarify the role of the stuck particles, we examine
how the motion of the mobile particles depends on their distance
to the nearest stuck particle.  It is important to
note that we only know the distance to the nearest {\it visible}
stuck particle, so some particles will appear far from any stuck
particle and yet be neighboring an invisible stuck particle.
With this caveat in mind, we plot the mean squared displacement
$\langle \Delta r^2 \rangle$ as a function of distance $s$ from
the nearest visible stuck particle in Fig.~\ref{stuckmobility}(a)
(using $\Delta t = 100$~s).  ($s$ is defined based on the initial
position of the mobile particle at $t$, rather than the final
position at $t+\Delta t$.)  Near a stuck particle, $\langle \Delta
r^2 \rangle$ decreases by about 10 - 20 percent.  The size of the
decrease appears about the same for all samples.  However, thicker
samples show a longer-ranged influence of the stuck particles;
we are not sure why.  For the $H=15.8$~$\mu$m data, the stuck
particle influence is seen even out to $s \approx 6$~$\mu$m, nearly
three small particle diameters.  Slight upturns are seen at
the smallest values of $s$ that we plot, around $s = 2.36$~$\mu$m
$=2 a_S$.  These are due to tracking errors (particles which
appear incorrectly closer to a stuck particle than their diameter),
which causes a larger subsequent displacement when the
tracking error is corrected.  For all curves in
Fig.~\ref{stuckmobility}(a), the data at large values of $s$
typically correspond to particles near the edge of our imaging
volume, where no visible stuck particles happen to be nearby
within the imaging volume.  Because of the high likelihood of
being adjacent to stuck particles outside the imaging volume
(given the fairly uniform coverage of the stuck particles on long
length scales), we crop the data for $s < 0.8H$ where we think
the data are potentially misleading.  In all cases, the cropped
data are essentially flat, that is, independent of $s$ and
continuing the trend shown in Fig.~\ref{stuckmobility}(a).

We note that the magnitude of the effect seen is not large, on the
scale of the dramatic slowing down shown in Fig.~\ref{avgmobility}.
That is, the slowing due to varying the sample thickness $H$ appears
to be the primary effect, and the influence of the texture is a
secondary effect.  For example, changing $H$ from 12.6~$\mu$m to
$8.06$~$\mu$m in Fig.~\ref{stuckmobility}(a) changes the overall
dynamics by a factor of more than 2, as indicated by the horizontal
dotted lines, while the proximity to stuck particles results in
a decrease by a factor of only 1.2 for each $H$.  The variation
between the amount of stuck particles on walls ranges from 10\%
to 20\% of the area, which our current results suggest is not
enough of a variation to change the overall character of the
slowed dynamics as a function of $H$ (in other words, the data
shown in Fig.~\ref{avgmobility}).

\begin{figure}
\begin{center}
\includegraphics[width=11.0cm]{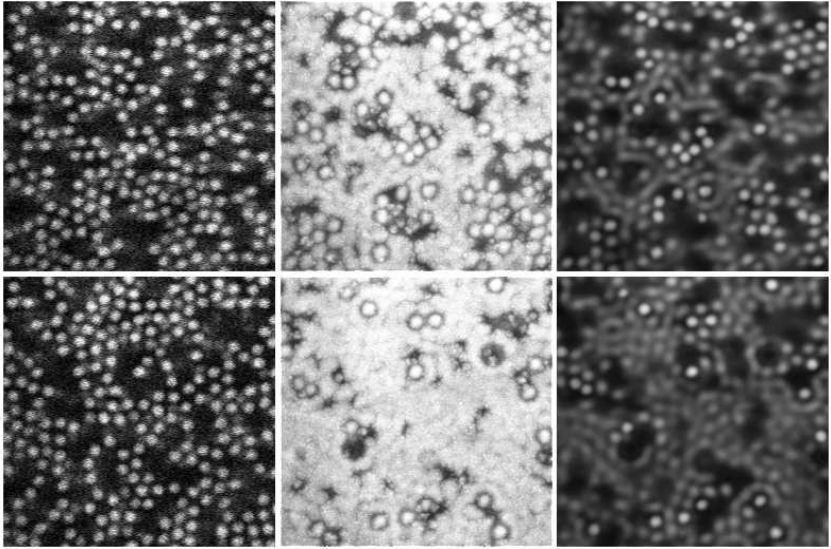}
\end{center}
\caption{
Images similar to those described in Fig.~\ref{montageone}.
Here, the top of the slide (top row of images) has more stuck
particles than the bottom of the slide (bottom row of images).
In particular, the bottom-middle image shows a large region where
mobile particles have wandered freely, whereas the top-middle image
shows much smaller free patches of open surface.  For this region
of the sample chamber, $H = 8.09$~$\mu$m.  The duration of this
movie was 2500~s.
}
\label{montagetwo}
\end{figure}

\begin{figure}
\centering
\includegraphics[width=8.5cm]{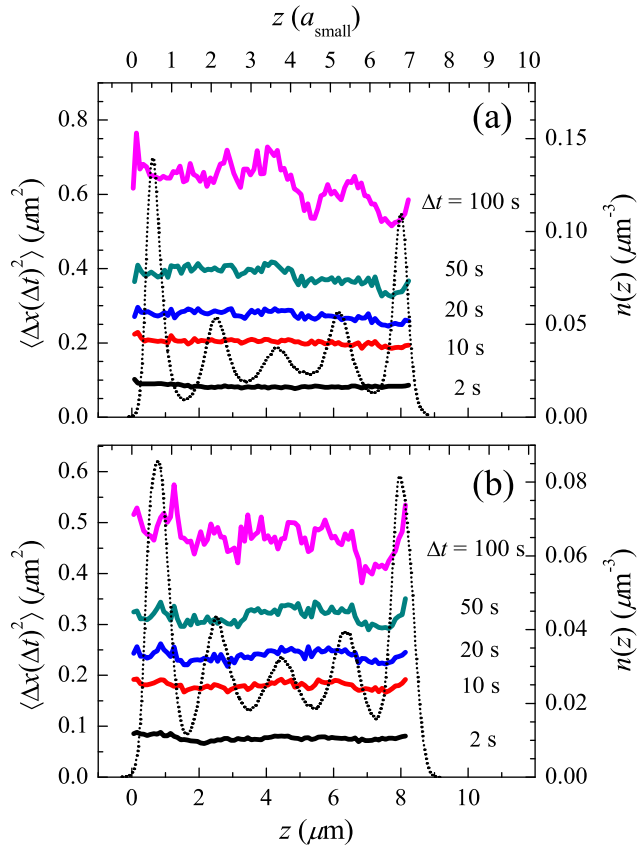}
\caption{
(Color online.)
In both panels, the thin solid black curve shows the local number
density $n(z)$ for the mobile particles (right vertical axis).  
The family of thicker
shaded curves show the average mobility as a function of $z$, for the
values of $\Delta t$ indicated.  For (a), the data
correspond to Fig.~\ref{montagetwo} with $H=8.09$~$\mu$m.  For (b),
the data correspond to Fig.~\ref{montageone} with $H=8.06$~$\mu$m.
The noise in the $\langle \Delta x^2 \rangle$ data 
give a good representation of the
uncertainty, $\pm 10$\% in all cases.
}
\label{nzmobile}
\end{figure}

Despite the weakness of the effect, these results suggest that
different boundary conditions will slightly influence the overall
dynamics.  For example, if there were more stuck particles on one
side of the sample chamber than on the other, we might expect a
gradient in mobility.  Such asymmetric boundary conditions are
present in a few of our experiments, such as the example shown
in Fig.~\ref{montagetwo}.  Here, the top of the sample chamber
has more stuck particles than the bottom.  This is most clearly
seen by contrasting the top-middle image (from the top
of the sample chamber) with the bottom-middle image.  To examine
the impact on the mobility of particles, we plot the average mean
square motion $\langle \Delta x^2 \rangle$ as a function of $z$
in Fig.~\ref{nzmobile}(a).  Here, the dotted black line indicates
the number density as a function of $z$, showing the layers of
particles, and the thicker shaded curves show the mobility
$\langle \Delta x^2 \rangle$ for different lag times $\Delta t$
as indicated.  Especially for the longest lag time $\Delta t =
100$~s, particles are more mobile closer to $z=0$, where there are
fewer stuck particles at the glass wall.  Figure \ref{nzmobile}(b)
shows the more common case that we observe, corresponding to
the more symmetric boundary conditions of Fig.~\ref{montageone},
and where there is little variation in mobility across the sample.

One last question is how the 
stuck particles influence the spatial
dynamical heterogeneity.  Prior work has shown that in
glassy materials, motion occurs intermittently in time
and in spatially localized regions at any given moment
\cite{kob97,donati98,sillescu99,ediger00}.  For example, microscopy
studies of dense colloidal suspensions found cooperative motion,
where groups of particles would all rearrange simultaneously
\cite{weeks00,courtland03,lynch08,marcus99,kegel00,konig05}.
Our results above show that proximity to stuck particles at
the boundary decreases the average mobility of particles; this
suggests that the most mobile particles undergoing cooperative
rearrangements are less likely to be near the stuck particles.


To check this, we define a highly mobile particle as one with
a displacement in the top 10 percentile of the displacement
distribution \cite{edmond10}.  If the sample behaves homogeneously,
we expect that in any given region, roughly 10\% of the particles
we observe there over time should be highly mobile.
We first define displacements using
the methods of Refs.~\cite{weeks00,donati98}:  $\delta r(t,\Delta t)
= {\rm max} | \vec{r}(t_i) - \vec{r}(t_j)|$, with $t \le t_i < t_j
\le t+\Delta t$; by allowing the time interval to fluctuate, this
removes some of the ``noise'' of the Brownian motion and highlights
the truly large motions.  We determine the threshold $\delta
r^*$ such that 10\% of the displacements are larger than this
threshold.  We then examine, as a function of $s$, what fraction
of particles at that distance have displacements larger than the
threshold.  The results are plotted in Fig.~\ref{stuckmobility}(b),
which would be a flat line at 10\% if there was no influence of
the stuck particles (indicated by the horizontal dashed line).
Instead, we see that the highly mobile particles are less
likely to be adjacent to stuck particles.  Clearly, the stuck
particles have a strong influence on the spatially heterogeneous
dynamics.  While the two thinnest data sets ($H=6.25$~$\mu$m
and $H=8.06$~$\mu$m, darkest curves) show a spike around $s =
3$~$\mu$m, it is most likely noise due to lack of statistics,
whereas the dip close to contact ($s \approx 2 a_S = 2.36$~$\mu$m)
is systematic for all data.  We estimate the uncertainties due to
noise in Fig.~\ref{stuckmobility}(b) are $\pm 2$\%, judging from
the large $s$ data.  The deviations for $s < 4$~$\mu$m appear
significant, especially the dip close to contact.

\section{Conclusions}

We have studied the influence of confinement on a dense colloidal
suspension, which is a good model system for a small-molecule
glass-former.  Confinement between two rigid glass walls results in
slower particle motion \cite{nugent07prl}.  In this article,
we examined how particles stuck on the walls influence the
dynamics of nearby particles.  Not surprisingly, proximity to a
stuck particle results in reduced mobility.  In cases where one
wall has more stuck particles than the opposite wall, a slight
gradient in mobility is seen, with the slowest motion next to
the roughest wall.  These results are in qualitative agreement
with simulations, although the roughness in simulations is of a
different character than our experiment \cite{kob02,lowen99}.

One caveat is that in our experiments, only the small particles
are fluorescently dyed and so we do not see the large particles.
There are both mobile large particles whose motion is unknown,
and also large particles stuck against the wall whose influence
is unseen.  Presumably these latter particles
diminish particle mobility nearby.  Currently, we see a reduction in
average particle motion of about 20\% near stuck visible particles.
Likely if we could measure the distance to the nearest particle
of either size, the observed reduction in average particle motion
near immobile particles would become more pronounced.  However,
it is possible that due to the layering, proximity to immobile
large particles would not slow down the mobile small particles.
Prior experimental studies of binary particle mixtures do not see
a strong overall difference in mobility of the small and large
particle species \cite{lynch08}, and our qualitative observations
of confined samples using DIC microscopy likewise did not see
large differences.

Our work shows that local variations in wall texture influence the
motion of nearby particles, and this in turn suggests that it is
possible to locally tune particle mobility by tuning wall texture.
While this observation is unsurprising, this
has implications for microfluidic flow of colloidal suspensions.
First, the effective viscosity of a colloidal suspension is likely
higher in a small capillary tube, given our observed glassiness
in the particle motion.  (However,
the viscosity and diffusivity may depend differently on the
confinement length scale \cite{mckenna05}.)  Second, if particles
stick to the capillary tube walls, then over time the mobility
will decrease especially strongly near the stuck particles.

We are currently conducting experiments to study more complex and
controlled wall textures to see their influence on the glassiness
of confined of colloidal suspensions, and we hope to report on
this work in the future.

This work was supported by a grant from the National Science
Foundation (DMR-0804174).  We thank A.~Schofield and
W.~C.~K.~Poon for providing our colloidal particles.

\bibliography{eric}

\end{document}